\documentclass[aps, pra, twocolumn, superscriptaddress, amsmath, 
tightenlines, longbibliography]{revtex4-2}
\usepackage{xcolor}
\usepackage{amssymb}
\usepackage{amsmath}
\usepackage{dcolumn}
\usepackage{graphicx}
\usepackage{mathrsfs}
\usepackage{appendix}
\usepackage{graphicx}
\usepackage{booktabs}
\usepackage{units}

\definecolor{Dred}{RGB}{190,0,0}
\def \hide#1{}

\newcommand{\ket}[1]{\mbox{$|#1\rangle$}}

\usepackage{url}
\usepackage[colorlinks]{hyperref}
\hypersetup{%
	plainpages=true,
	breaklinks=true,       
	hypertexnames=false,  
	pageanchor=true,
	colorlinks=true,
	linkcolor={blue},
	citecolor={red},
	urlcolor={blue},
	anchorcolor={black}
}

\begin{document}

\title{Realizing quantum optics in structured environments with giant atoms}

\author{Xin Wang}
\affiliation{Institute of Theoretical Physics, School of Physics, Xi'an Jiaotong University, 
Xi'an 710049, People’s Republic of China}

\author{Huai-Bing Zhu}
\affiliation{Institute of Theoretical Physics, School of Physics, Xi'an Jiaotong University, 
	Xi'an 710049, People’s Republic of China}

\author{Tao Liu}
\affiliation{School of Physics and Optoelectronics, South China University of Technology,  
Guangzhou 510640, China}

\author{Franco Nori}
\affiliation{RIKEN Center for Quantum Computing, Wakoshi, Saitama 351-0198, Japan}
\affiliation{Theoretical Quantum Physics Laboratory, RIKEN Cluster for Pioneering Research, 
Wako-shi, Saitama 351-0198, Japan}
\affiliation{Physics Department, The University of Michigan, Ann Arbor, Michigan 48109-1040, USA}
\date{\today}

\begin{abstract}
To go beyond quantum optics in free-space setups, atom-light interfaces 
with structured photonic environments are often employed to realize unconventional quantum 
electrodynamics (QED) phenomena. However, when employed as quantum buses, those long-distance 
nanostructures are limited by fabrication disorders.
In this work, we alternatively propose to realize structured light-matter interactions by  
engineering multiple coupling points of hybrid giant atom-conventional environments without  
any periodic structure. We 
present a 
general optimization method to obtain the real-space coupling sequence for 
multiple coupling points.
We report a broadband chiral emission for 
frequency-tunable giant emitters, with no analog in 
other quantum setups. Moreover, we show that the QED phenomena in the band 
gap environment, such as fractional atomic decay and dipole-dipole interactions mediated by a 
bound state, can 
be observed in our setup. Numerical 
results indicate that our proposal is robust against fabrication disorders of the
coupling
sequence. Our work opens up a new route for realizing unconventional light-matter interactions.

\end{abstract}

\maketitle

\section{Introduction}
Harnessing interactions between quantum emitters and quantized 
electromagnetic fields is a central topic
of quantum 
optics~\cite{Scully1997,Agarwal2012,Kimble1977,cohen1998atom,Gu2017,Kockum2019b_r}.
In recent years, a burgeoning paradigm with
giant atoms, which are coupled to waveguides at multiple separate points with their sizes 
comparable to photonic wavelengths, 
provides unanticipated opportunities to gain insights into exotic quantum 
optics beyond the dipole approximation~\cite{Gustafsson2014,Kockum2014,Andersson2019,
Zhao2020,Guo2020,Du2021,Soro2021,Yin2022,Du2022L,Wang2022,Xiao2022,Cheng2022,Qiu2023,Jia2023,Santos2023}.
The nonlocal coupling points cause  nontrivial phases accumulation of the 
propagating field from a single giant emitter~\cite{Kannan2020,Guo2017,Wang2021},   allowing 
to observe exotic phenomena with no analog in small atom setups. The examples include 
decoherence-free 
interaction and oscillating bound states, which are caused by quantum interference 
and time-delay effects, respectively~\cite{Kockum2018,Guo2019,Lim2022}.

Structured dielectric environments, which are scalable in 
integrated chips, have achieved tremendous 
progresses in quantum 
optics~\cite{Lodahl2015,Goban2015L,Roy2017,Chang2018,Yu2019,Scigliuzzo2022,Tang2022}.
Compared with free-space setups,
the vacuum mode properties and dispersion relation can be 
tailored freely by 
shaping the dielectric 
structures~\cite{Kofman1994,Smith2000,Lambropoulos2000,Kafesaki2007,Lu2014,Indrajeet2020,Stewart2022}.
One emblematic example is 
photonic crystal waveguides (PCWs), where the
dielectric profile is 
periodically modulated, leading to the appearance of band 
gaps~\cite{John1990,Kien2004,Kien2005,Vetsch2010,Hung2013,Corzo2016,Liu2017}.
Inside the gap, stable bound states of the hybrid photon and emitter are formed, which can 
alternatively
mediate long-range interactions between 
emitters~\cite{Zhou2008L,Liao2010,Goban2014,Douglas2015,Douglas2016,Hood2016,GonzlezTudela2015,Munro2017,Leonforte2021}.
Moreover, when light is tightly transversely confined in high refractive-index 
materials, chiral emission occurs in nanophotonic structures 
owing to spin-momentum locking~\cite{Mitsch2014,Bliokh2015, Petersen2014, Young2015, 
Sllner2015,Bliokh2015b,Coles2016,Lodahl2017}. 
However, fabricating  long-distance  nanostructures is very 
challenging when configuring those nanophotonic materials 
as quantum buses for quantum information processing~\cite{Goban2014,Douglas2015}. 
Due to unavoidable fabrication
disorders and defects, photons are scattered repeatedly, and their
fragile quantum coherence is
destroyed~\cite{Patterson2009,Garc2010,Lang2015,Mann2015}. 
 
Here we show that structured light-matter 
	interactions can be realized from the viewpoint of giant 
	atoms, i.e.,
	by spatially designing the
	coupling sequence with a conventional photonic waveguide without any 
	periodic 
	structure. We 
	present a 
	general optimization method to
	obtain real-space coupling sequences for a target structured 
	environment, which has 
	never been discussed in previous studies.
	As examples, we show that both
	broadband chiral emission for 
	frequency-tunable giant emitters (with no analog in 
	other quantum setups) and band-gap effects	can be realized by considering tens 
	of coupling points in a conventional one-dimensional (1D) waveguide.
Numerical results indicate that our proposal is robust against
fabrication disorders in the coupling sequences, and can avoid localization and decoherence 
of photons appearing in long-distance nanostructures.

\begin{figure}[tbph]
	\centering \includegraphics[width=8.7cm]{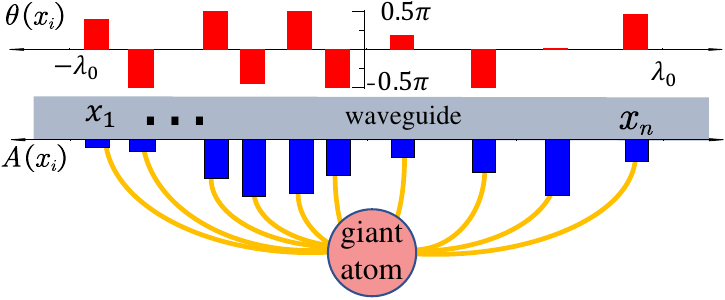}
	\caption{(a) Sketch of our proposal: A giant atom couples to a 
		conventional 1D waveguide at positions $x_1,...,x_N$. The coupling 
		sequence 
		$g(x_i)=A(x_i)e^{i\theta(x_i)}$ is obtained via optimization 
		methods.}
	\label{fig1m}
\end{figure}

\section{Optimizing coupling sequence}
The generic Hamiltonian of a quantum emitter interacting 
with a bosonic bath can be written as (setting $\hbar=1$)
\begin{equation}
	H_{\mathrm{int}}=\sum_k{\Delta _ka_{k}^{\dagger}a_k}+\sum_k{G_k\left( 
		a_{k}^{\dagger}\sigma _-+\mathrm{H}.\mathrm{c}. \right)},
	\label{Hrtot}
\end{equation}
where $\Delta _k=\omega_k-\omega_q$, with $\omega_q$ being the atomic transition 
frequency. Assuming a giant atom interacting with the waveguide at multiple points 
$X=\{x_1,...,x_N\}$ 
(see 
Fig.~\ref{fig1m}), the $k$-space interaction is thus written as
$G_k =\sum_{x_i}{g\left( x_i \right) e^{-ikx_i}}$, with $g(x_i)=A(x_i)e^{i\theta(x_i)}$ being 
the
interaction strength at $x_i$ (see  
Fig.~\ref{fig1m}). 
For small-atom setups, $G_k$ is approximately a constant due to the point-like coupling 
between 
the emitter and waveguide. Therefore, the structural engineering of the photonic waveguide's
dispersion relation $\Delta_k$ plays an important role in achieving exotic quantum 
dynamics in previous studies~\cite{Zhou2008,GonzlezTudela2017,Tudela2017L,Bello2019,Kim2021}. 
In contrast, for our proposal,  the bosonic environment is no longer designed, and a 
\textit{conventional waveguide} is used. It has a linearized dispersion within the photonic 
bandwidth to 
which the giant atom significantly couples, i.e., $\omega_k=c|k|$ with $c$ being the group 
velocity.
An intuitive method for realizing the desired $G_k$ is to 
find the real-space function $g(x_i)$ via inverse Fourier transformation 
(iFT)~\cite{Kockum2014}. However, our following discussions indicate that 
this method has many problems and introduces many experimental 
overheads.


\subsection{Analytical method and its problems}
We assume giant atoms interacting with a 1D waveguide, which has a 
linearized dispersion within the 
photonic bandwidth to which the giant atom significantly couple. 
The following $k$-space 
coupling function equivalently 
describes an atom interacting with a band-gap
environment
\begin{equation}
	G^I_k =\begin{cases}
		0 \qquad  \left\{ k_0-\frac{k_d}{2}<|k|<k_0+\frac{k_d}{2}\right\},\\
		G_0       \qquad       
		\left\{k_0+\frac{k_d}{2}<|k|<k_{\text{max}}\right\}, \\
		G_0 \qquad  
		\left\{-k_0+\frac{k_d}{2}<k<k_0-\frac{k_d}{2}\right\}.\\
	\end{cases}
\end{equation}
That is, the gap's width is $k_d$ and is centred at $\pm k_0$. 
The coupling constant is denoted by $G_0$. For convenience the ultraviolet 
cut-off
frequency 
is set at $ck_{\text{max}}$, which should be large enough 
to approximate the regime 
$\left\{k_0+\frac{k_d}{2}<|k|<k_{\text{max}}\right\}$ 
as an infinite 
bandwidth environment.
\begin{figure*}[tbph]
	\centering \includegraphics[width=17cm]{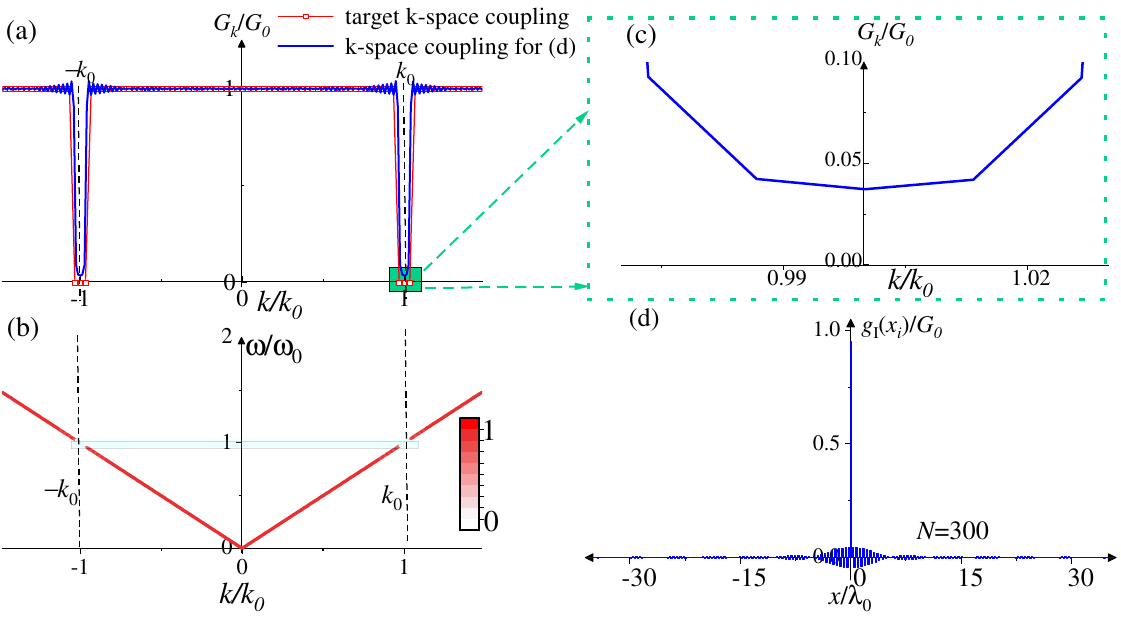}
	\caption{(a) The $k$-space coupling for the real-space 
		sequence by discretizing $g_I\left( x \right)$ in Eq.~(\ref{GIa}) 
		[see plot in (d)]. 
		The target coupling function has two symmetric 
		dips centred at $k_0$ with width $k_d/k_0=1/15$. (b) The waveguide 
		is 
		assumed to be of linear dispersion with a phase \textsf{velocity} 
		$c$. The 
		coupling strength $G_k$ in (a) is mapped with color, where two 
		symmetric 
		dips around $k_0$
		are equivalent to band gaps in a structured environment. (c) The 
		enlarged plot  around $k_0$ of plot (a). Inside the band gap there is 
		still 
		remnant 
		non-zero coupling ($\sim 0.02G_0$) even with a large sampling number 
		$N=300$. 
		(d) Real-space coupling sequence $g(x_i)$ (with $N=300$). 
		The sampling interval and total length are set as
		$X_T=0.24\lambda_0$ and $L=35\lambda_0$, respectively.}
	\label{fig2m}
\end{figure*}

The inverse Fourier transform (iFT) of $G^I_k$ is derived as
\begin{equation}
	g_I\left( x \right) =\frac{\sin k_{\max}x}{\pi x}-2\frac{\sin k_dx/2}{\pi 
		x}\cos 
	k_0x, \label{GIa}
\end{equation}
which is a continuous function in real space. In experiments, giant atoms 
usually couple at multiple discretized positions on a waveguide.
Therefore, we assume that the coupling function 
$g_I\left( x \right)$ is sampled by the 
following function 
\begin{eqnarray}
	S(x)&=&W\left( x \right)P(x), \quad
	W\left( x \right) =\left\{ \begin{array}{c}
		1  \quad  |x|\leq L,\\
		0  \quad   |x|>L,\\
	\end{array} \right. \notag \\
	P(x)&=&\sum_{n=-\infty}^{n=+\infty}\delta \left( x-nX_T \right),
\end{eqnarray}
where $W\left( x \right)$ is a window function with a width $2L$, and 
$P(x)$ is the sample sequence composed by $\delta$-function series which 
are equally spaced with distance $X_T$. To avoid retardation 
effects, the total coupling length $2L$ (i.e., the giant atom's size) should 
be much smaller 
than the size of the decaying photonic wavepacket~\cite{Guo2019}. All those 
physical restraints 
will be addressed in later discussions.

According to the convolution theorem, the Fourier transformation of $S(x)$ is 
written as 
\begin{eqnarray}
	S(k)&=&W\left( k \right)*\Delta(k), \quad W\left( k \right)=
	2\frac{\sin(kL)}{k}, \notag \\
	P(k)&=&\frac{2\pi}{X_T}\sum_{n=-\infty}^{n=\infty}\delta(k-\frac{2\pi 
		n}{X_T}),
	\label{Sk}
\end{eqnarray}
where $*$ represents the convolution of two functions. Equation~(\ref{Sk}) 
indicates that the width 
of the $\delta$-functions $\delta(k-\frac{2\pi n}{X_T})$ in $k$-space 
are broadened as $\sim 2\pi/L$. To resolve 
the narrow band gap, the following relation should be satisfied 
\begin{equation}
	2\pi/L\ll k_d \, \,\longrightarrow L \,\, \gg \frac{2\pi}{k_d}.
\end{equation}
Additionally, to avoid spectrum aliasing effect,
the sample distance is bounded by the Nyquist–Shannon sampling theorem
\begin{equation} 
	\frac{2\pi}{X_T}\gg 2k_{\text{max}} \, \,\longrightarrow  \, \, X_T\ll 
	\frac{\pi}{k_{\text{max}}}.
\end{equation}
Consequently, the coupling number of the giant atom is bounded by
\begin{equation}
	N=\frac{2L}{X_T}\gg \frac{4k_{\text{max}}}{k_d}.
	\label{Nbound}
\end{equation}

We now consider the target 
$k$-space coupling function with $k_{\text{max}}=2k_0$ and $k_d=1/15$ 
[see 
Fig.~\ref{fig2m}(a)]. According to Eq.~(\ref{Nbound}),
the minimum coupling number is calculated as $N=120$. In 
Fig.~\ref{fig2m}(d), we plot the sampled real-space
coupling sequence by setting $N=300$. The corresponding $k$-space coupling 
function is shown in 
Fig.~\ref{fig2m}(a), and the amplitude is mapped with color in 
Fig.~\ref{fig2m}(b). Note that $\lambda_0=2\pi/k_0$ is the wavelength of the
central 
mode in the gap, and 
is employed as the unit length in this work. The unit for the frequenc is 
adopted as $\omega_c=ck_0$. To mimic the band gap, 
the 
most 
important feature of 
$G_k$ is the vanishing of the
coupling strength around $k_0$. The enlarged plot 
of this regime is depicted in Fig.~\ref{fig2m}(c), which shows that the 
remnant 
coupling is still about $0.02G_0$. 
In principle one can keep increasing both $N$ and $L$ to 
suppress the 
non-zero coupling. However, much more coupling points are need, which is very 
challenging in experiments. Moreover, given that $L$ is comparable to the 
wavepacket 
decaying from a single point, the propagating time cannot be neglected.

\begin{figure}[tbph]
	\centering \includegraphics[width=8.9cm]{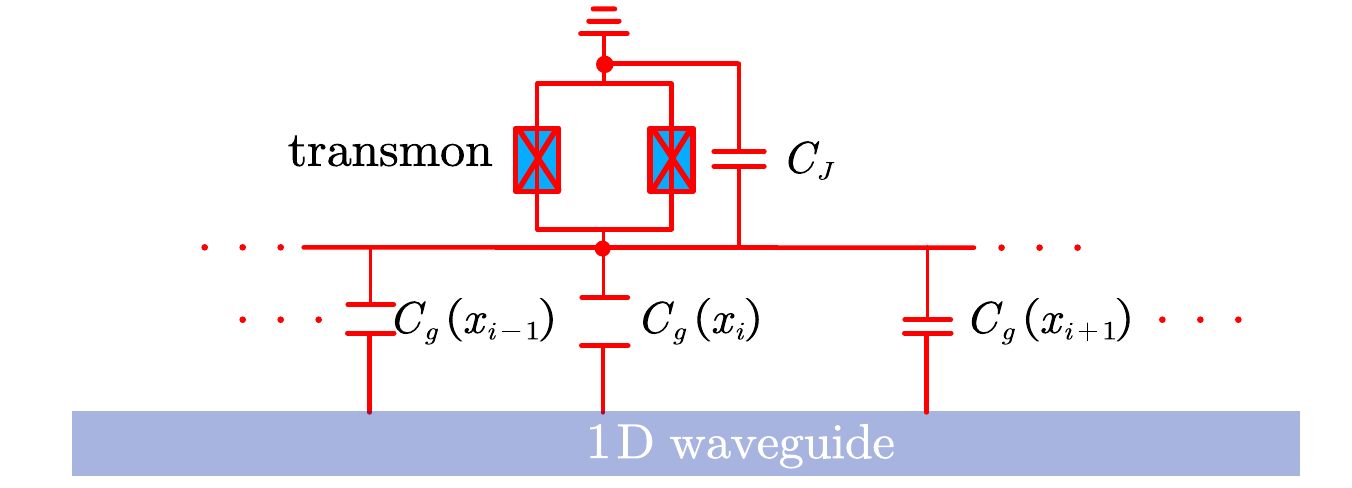}
	\caption{Sketch of a feasible setup to realize our proposal: a 
		transmon couples to a 1D coplanar 
		waveguide at multiple coupling points via local capacitances 
		$C_g(x_i)$. The real-space 
		discretized coupling 
		function is encoded into the capacitance sequence $C_g(x_i)$.}
	\label{fig3m}
\end{figure}

Additionally, we note that the coupling strengths $g(x_i)$ which are sampled 
from 
Eq.~(\ref{GIa}), alter 
their signs [see Fig.~\ref{fig2m}(d)]. That is, $g(x_i)$ needs an additional 
$\pi$-phase difference, which leads to 
another problem when implementing the coupling sequence in experiments. We 
now consider the circuit-QED as an
example, where giant atoms are mostly discussed. As depicted in 
Fig.~\ref{fig3m},
a transmon (working as a giant atom) is capacitively 
coupled to a 1D waveguide at multiple points. As discussed in 
Ref.~\cite{Koch07,Gu2017}, the interaction strength
is 
written as
\begin{gather}
	G_k = \sum_{x_i} g(x_i) e^{-i k x_i}, \\
	g(x_i)= \frac{e}{\hbar} \frac{C_g(x_i)}{C_\Sigma}\sqrt{\frac{\hbar 
			\omega_k}{C_t}}\simeq \frac{e}{\hbar} 
	\frac{C_g(x_i)}{C_\Sigma}\sqrt{\frac{\hbar 
			\omega_q}{C_t}}, \label{g_real_space} 
\end{gather} 
where $C_J$ is the Josephson
capacitance of the transmon,  $C_g(x_i)$ is the coupling capacitance at point 
$x_i$, $C_\Sigma=C_J+\sum_i C_g(x_i)$, and $C_t$ is the total capacitance of 
the 
waveguide.
In Eq.~(\ref{g_real_space}) we replace $\omega_k\rightarrow \omega_q$
for the zero-point fluctuations of the voltage operators because only modes 
around $\omega_q$ contribute significantly to the dynamics. Under this 
condition, the 
local coupling 
strength $g(x_i)$ is proportional to the coupling capacitance $C_g(x_i)$, and 
the coupling 
sequence in Fig.~\ref{fig2m}(d) 
can be directly encoded into $C_g(x_i)$ (see Fig.~\ref{fig3m}). We notice 
that 
the coupling signs of $g(x_i)$ are
fixed because $\{C_{g}(x_i),C_{\Sigma}\} \geq 0$. That is, there are \emph{no 
	additional} $\pi$-phase differences between different coupling points, 
and the 
discretized 
coupling obtained via the iFT method cannot be implemented with a linear 
coupling 
capacitance (or inductance). The additional local phase $\theta(x_i)$ can be 
encoded at $x_i$ 
via the
time-dependent modulating of the nonlinear QED elements, which however will 
add 
more overheads in the
experiments~\cite{Chen2014,Wulschner2016,McKay2016},

In conclusion, to realize a structured environment
with a giant atom, the analytical iFT method has the following problems: 

1) Too many coupling points might be needed, which is challenging for the 
experimental 
realization.

2) The remnant non-zero coupling in the band gaps is still high.

3) The coupling strengths alter their signs, which is unfeasible with the 
linear 
coupling elements used in the experiments.

\subsection{Optimization method}

To solve the above problems, we now present a general optimization algorithm 
to find the desired coupling sequence.
Unlike previous setups using identical giant atom-photon interacting strength 
at each coupling 
point and equal distances between coupling 
points~\cite{Kockum2018,Guo2019,Lim2022}, our 
device relies on the optimal design of the coupling sequence.

We consider the unequal 
contribution of the modes with different unbalanced weights. For the band-gap 
environment, the 
coupling strength should be exactly zero inside the gap area, which is the 
most 
important feature for a band-gap environment. 
Outside the 
band gap area, even if the interaction varies with $k$ slightly (of the same 
order), the dynamics, such as trapped bound state and non-exponential decay 
led by band 
gaps, can still be observed. For the modes far away from the gap, strong 
coupling strengths does not affect the system's evolution due to large 
detuning relations. 
Therefore, the constraint requiring the $k$-space coupling 
	strength outside the band gap to be identical, is too strong.
All these indicate that the desired $k$-space coupling can 
be obtained even when the real-space coupling strengths are of the 
same sign. Moreover, relaxing the restrict conditions by allowing $G_k$ to 
vary
with $k$ can 
also reduce the required number of coupling points.

Now we convert realizing the target $k$-space coupling function 
as an optimization problem, which can be
solved numerically. Given that the QED setup is constructed via conventional 
linear elements, the 
local phase $\theta(x_i)\equiv0$, and
$G^*_k=G_{-k}$ is valid.  
In this case,  $\forall g(x_i) \geq 
0$ should be added in the constraint conditions. The constraint conditions 
for this problems are 
summarized as follows:
\begin{eqnarray}
	\begin{cases}
		1.\qquad  \forall g(x_i) \geq 0, \\
		2. \qquad  \eta\lambda _0<\min \left\{ x_{i+1}-x_i \right\},\\
		3. \qquad  -\frac{L}{2}<x_1<x_N<\frac{L}{2}, \quad L\ll \bar{L}_0, 
		\quad   \\
		4. \qquad  N\leq N_{\text{max}},\\
	\end{cases}\label{constraint}
\end{eqnarray}
where $\bar{L}_0=(\sum_{i}^{N} L_i)/N$ with  $L_i$ being the size of a 
decaying photonic wavepacket from a 
small
atom which just couples to the waveguide at a single point $g(x_i)$.
Condition 1 restricts that all real-space coupling strengths are of the same 
sign, which avoids the coupling sign problem in QED setups with linear 
couplers. 

Condition 2 sets the lower bound of the distance between two
neighbor points.
The reason for this restriction is that the coupling is mediated via physical 
elements with finite sizes (for example, 
capacitances or inductances in circuit-QED). Due to fabrication
limitation and to avoid crosstalk, two neighbouring points cannot be too 
close to 
each other.

In condition 3, $\bar{L}_0$ is the average size of all the decaying 
wavepackets. This restriction guarantees that the re-absorption and 
re-emission of photons due to time retardation can be neglected. 

Condition 4 
sets the maximum coupling 
number, which is much smaller than that bounded by Eq.~(\ref{Nbound}).

\begin{figure*}[tbph]
	\centering \includegraphics[width=15cm]{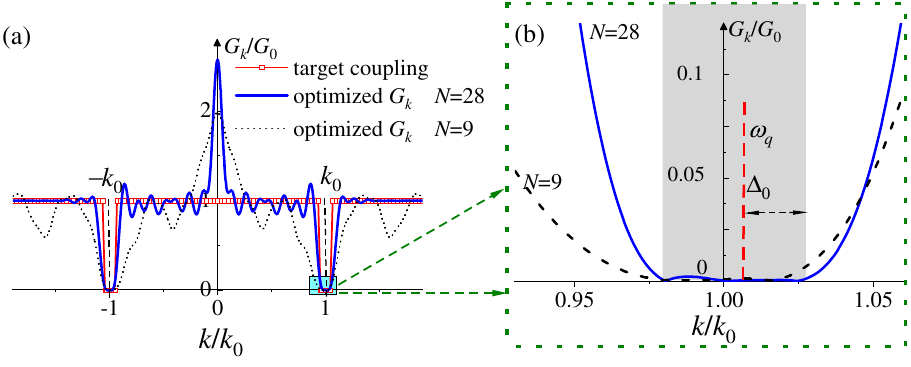}
	\caption{(a) Coupling 
			function $G_k$ obtained via out
			optimization 
			algorithm with $N=28$ and $N=9$. The real-space coupling 
			sequence for 
			$N=28$ is listed in Table~I of Appendix A. (b) The enlarged plot 
			around 
			$k_0$. The detuning between $\omega_q$ and the band edge of $N=28$
			is 
			denoted as $\Delta_0$.
	}
	\label{fig4m}
\end{figure*}

Considering a real-space sequence $g'(x_i)$ 
satisfying condition (1-4), its $k$-space coupling function is denoted as 
$G'(k)$, 
which is obtained from Eq.~(\ref{g_real_space}). 
To find the optimized real-space sequence,
we define an 
objective function 
\begin{equation}
	C_m=\int_{-k_\text{max}}^{k_\text{max}}dk||G'(k)|-|G^I(k)||w(k),
	\label{C_ms}
\end{equation}
which can quantify the difference between obtained $G'(k)$ and the target 
coupling 
function. In Eq.~(\ref{C_ms}) we introduce a weight function $w(k)$ to 
control 
the similarities
for modes in different regime regime. For simplicity, in this work we assume 
$w(k)$ to be 
\begin{equation}
	w(k)=\begin{cases}
		w_1 \qquad  \left\{ k_0-\frac{k_d}{2}<|k|<k_0+\frac{k_d}{2}\right\},\\
		w_0       \qquad       
		\left\{k_0+\frac{k_d}{2}<|k|<k_{\text{max}}\right\}, \\
		w_0      \qquad  
		\left\{-k_0+\frac{k_d}{2}<k<k_0-\frac{k_d}{2}\right\}. \\
	\end{cases}
\end{equation}
Since the similarity between $G'(k)$ and $G^I(k)$ in the band gap regime is 
much more important, we set $w_1\gg w_0$ in Eq.~(\ref{C_ms}). The 
	optimization 
	process minimizes $C_m$ by searching the possible functions $g(x_i)$ 
	satisfying the constraint condition in Eq.~(\ref{constraint}).
Note that the constraint conditions stated in 
Eq.~(\ref{constraint}) can be 
different, depending on problems studied and experimental setups employed.

To simulate a band-gap environment, we set  $k_d/k_0=1/15$,
$L\simeq 17\lambda_0$, $\eta=0.1\lambda_0$, 
$N_{\text{max}}=30$, and
$w_1=60w_0$, and the obtained
coupling sequence 
$g(x_i)$ (of the same sign) is listed in Table~\ref{table1} of Appendix A. 
We employ the proposed optimization
method  by searching the sets
$\{x_i, g(x_i)\}$. The optimized $G_k$ is depicted in 
Fig.~\ref{fig4m}(a), and the enlarged plot around the 
band 
gap is in Fig.~\ref{fig4m}(b). 
We find that the number of points is reduced 
as 
$N=28$, 
and the remnant non-zero coupling in gap area is decreased below 
$10^{-4}G_0$, 
which is much weaker than those in Fig.~\ref{fig2m}(c). 
Specially, $G_k$ varies slightly with $k$ for the modes outside the band gap, 
and 
the dc part (around $k\simeq0$) 
will strongly couple to the giant atom because the $g(x_i)$'s signs are the 
same [see 
Fig.~\ref{fig4m}(a)]. To 
demonstrate the band-gap effect, the atomic frequency 
is usually set around $ck_0$, and 
therefore, the interaction with those low-frequency components is negligible 
due 
to large detuning effects, which can be verified from the numerical 
discussion 
in next section.

In experiments, reducing the number of coupling points can 
	significantly simplify the whole setup. However, when $N$ is too 
	small, $G'(k)$ obtained by the proposed optimization algorithm 
	will differ significantly from the target coupling function $G_k$. For 
	example, when 
	$N=9$, 
	we find $G'(k)$  unable to capture the important 
	features of the band-gap environment, as 
	shown in Fig.~\ref{fig4m}. Outside the band-gap area, the 
	interaction strength begins to vary drastically. Moreover, the  
	band gap obtained becomes much wider than the target coupling $G(k)$. 
	Therefore, there 
	is a lower bound for the coupling number in our proposal, below 
	which the proposed algorithm cannot successfully find a suitable 
	sequence.

We now summarize non-trivial 
differences between 
interferences  
in the photonic structure and giant atoms. 
First, photonic media are often fabricated with periodic structures to 
satisfy the Bragg reflection relation. However, in our proposal with giant 
atoms, 
the 
coupling points can distribute with unequal 
spacing, and the 
coupling 
amplitudes can also differ considerably (see Table~	I of Appendix A).

Second, in principle periodic photonic structures should be infinitely long 
(or at least much larger than the wavelength/wavepacket size); 
otherwise the photons will be reflected by the boundary and the properties 
of structured environments cannot be observed.  When configured as a quantum 
bus, the quantum 
coherence will inherently be destroyed by fabrication disorders 
along the long waveguide. For proposals with giant atoms, the coupling 
sequences 
are
of finite length, with a finite number of couplings. 

Those differences indicate that the 
interference mechanisms in those two paradigms are fundamentally different, 
even 
though the 
observed
quantum optical phenomena are similar.
Revealing the interference mechanism in 
giant 
atoms is of both fundamental and technological significance, which cannot
follow the old routines used in 
photonic media.
\section{ QED phenomena in band-gap environments}
\subsection{Bound states}  
\begin{figure*}[tbph]
	\centering \includegraphics[width=18cm]{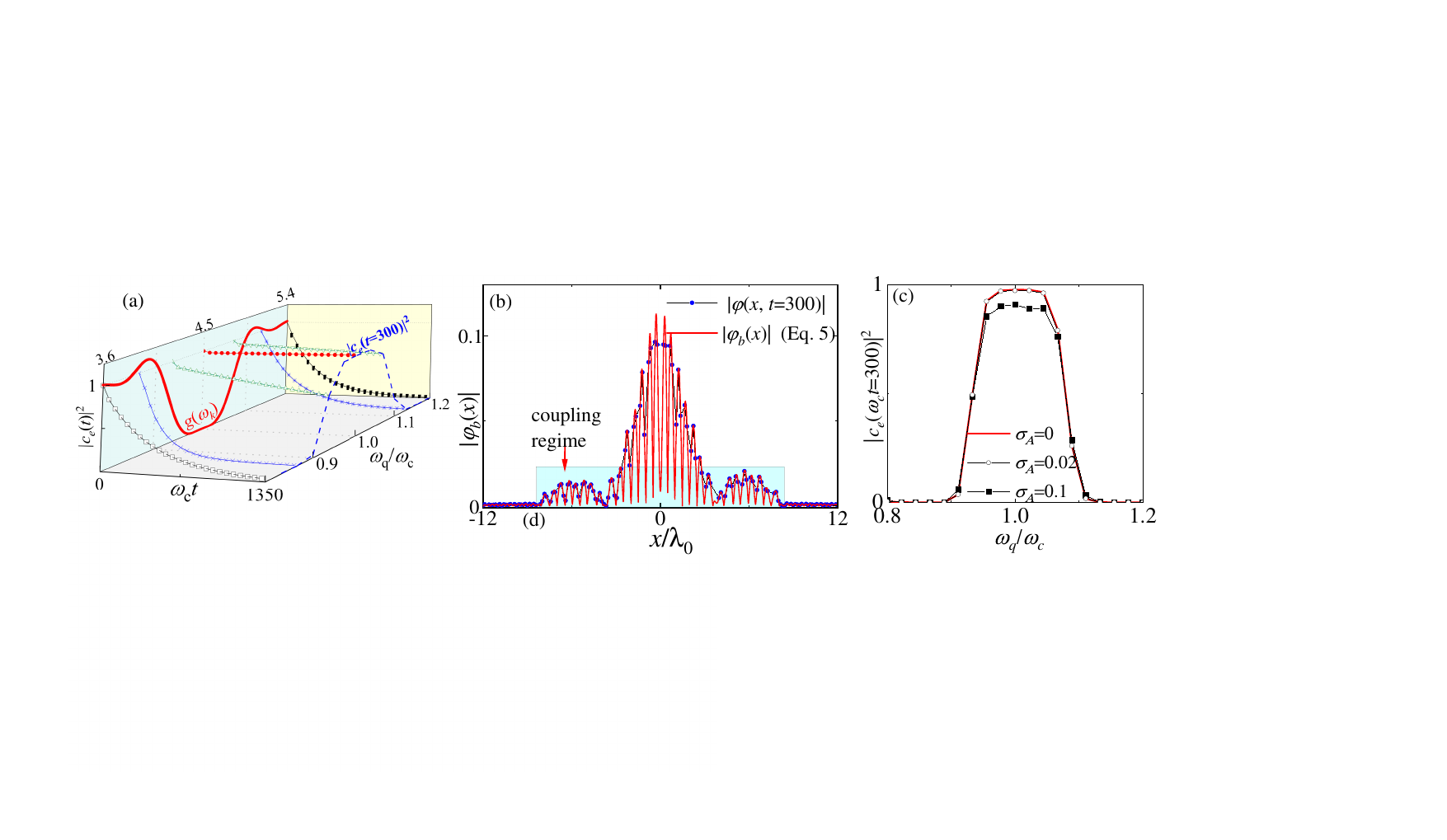}
	\caption{(a) Time evolution of $|c_e(t)|^2$ for 
		different atomic frequency $\omega_q$. (b)
		The photonic part of the bound state
		by setting $\omega_q=\omega_c$ (in the band gap). 
		(c) In the present of different disorder strengths, the trapped 
		population 
		$|c_e(t)|^2$ ($\omega_c t=1350$) changes with $\omega_q$.}
	\label{fig5m}
\end{figure*}
In Fig.~\ref{fig4m}(a), $G_k$ is zero only in a 
narrow band with width $k_d$ around $\pm k_0$, otherwise it remains a 
constant. This scenario is very similar to an atom 
interacting with 
a waveguide environment with band gaps 
in periodic structures. We now show that our setup shares very similar QED 
phenomena with conventional 
light-matter hybrid structures with photonic band gaps. Owing to 
suppression effects of the unbalanced weight function, the remnant coupling 
in the gap area is 
approximately zero, which can exactly mimic a band-gap environment.

We mainly focus on the fractional 
decay 
and bound state of the setup. 
In the single-excitation subspace, considering an initial excitation in the 
giant atom, the time-dependent state vector of the hybrid system is
$|\psi(t)\rangle=\sum_{k} 
c_k(t)|g,1_k\rangle+c_{e}(t)|e,0\rangle$. The evolutions of the atomic population $|c_e(t)|^2$ 
are 
shown in Fig.~\ref{fig5m}(a) for different $\omega_q$. There, $|c_e(t)|^2$ 
shows
fractional decay with most energy 
being trapped inside the atom when $\omega_q$ is in the band gap. 
As discussed in Appendix A, the trapped population is 
approximately 
$|c_e(t\rightarrow\infty)|^2\simeq (1+\sum_k{\left| 
	G_k/\Delta_k\right|^2})^{-2}$.
Once $\omega_q$ is shifted far away from the gap area, 
$|c_e(t)|^2$ can exponentially decay to zero. 
Moreover, there exists a
static bound state with its wavefunction localized around the atom. As 
derived in 
Appendix A, the real-space
distribution for the photonic part of the bound state is
\begin{equation}
	\psi_b(x)  
	\propto
	\int_{\infty}^{\infty} dk\frac{G_k}{\Delta_k}e^{kx}=\sum_{i=1}^N\int_{-\infty}^{\infty}  
	\frac{
		g(x_i) } {\Delta_k}e^{k(x-x_i)}dk.
	\label{phi_bx_ms}
\end{equation}

In Fig.~\ref{fig5m}(b), we plot the 
field distribution by solving the system evolution to $\omega_c t=1350$, 
which 
is well 
described by the stable bound state obtained 
in Eq.~(\ref{phi_bx_ms}).
All the above phenomena are very similar 
to those observed in setups with band-gap 
environments~\cite{Goban2014,Douglas2015,Douglas2016,Hood2016,GonzlezTudela2015}.
The counterintuitive phenomenon is that
\emph{there is no stable bound state if a small atom is coupled to the conventional 
waveguide.} 
While, for giant atoms coupled to the waveguide, fields emitted from different coupling points 
 interfere with each other [see Eq.~(\ref{phi_bx_ms})], which results in a  
time-independent $\phi_b(x)$. 
Note that, in our discussion, the propagating 
time inside the giant atom is negligible. 
Since the waveguide supports only modes with nonzero group 
velocity, the wavepacket outside the coupling regime 
cannot be reflected by any point and
will propagate away. Therefore,  $\phi_b(x)$ exactly lies 
within
	the coupling regime, which can be viewed as  bound states in continuum 
	studied in Ref.~\cite{Guo2019,Guo2020}.  The structured environment 
	supports 
	modes with
zero group 
velocity~\cite{Douglas2015,GonzlezTudela2015,Bello2019}, $\phi_b(x)$ can spread 
far away from the coupling 
point.

To include disorder effects, we sample 
the error $\delta\! g(x_i)$ randomly from a 
Gaussian distribution centered around zero with width $\sigma_A
g(x_i)$. The simulation method is presented in Appendix B.
We plot the disorder averaged excitation being trapped 
inside the atom versus $\omega_q$ ($\omega_c t=1350$), as shown in 
Fig.~\ref{fig5m}(c). 
The averaged coupling inside the gap becomes 
non-zero due to the random noise. Therefore, the protection from the band gap 
 is destroyed, 
and   $|c_e(t)|^2$ decays. The decoherence rate increases with growing $\sigma_A$, as shown in
Fig.~\ref{fig5m}(c). 
However, for 
$\sigma_A<0.1$, the evolution is only slightly affected by disorders. The 
experiments with giant emitters (with two or three coupling 
	points) in Refs.~\cite{Kannan2020,Kannan2023} indicate that the 
	amplitudes of the coupling 
	sequence can be fabricated with a high accuracy. There is no fundamental 
	limitation of increasing the coupling number to be 
	tens 
	of couplers. 
	With the 
	development of fabrication method, we believe that our
	proposal is within the capability of setups
	with giant atoms in the near future.

\subsection{Dipole-dipole interactions}
Considering multiple giant atoms coupled to a common waveguide with the 
optimized 
sequences for band-gap environments, these will interact each 
other given that their bound states overlap with each other. 
We derive their dipole-dipole interaction 
strength by taking two giant atoms as an example. The Hamiltonian describing 
two giant atoms 
interacting with a 
common waveguide is expressed as
\begin{eqnarray}
	H_{\text{int2}} =\!\! \sum_{k} \Delta_k a^\dag_k a_k
	+\! \sum_{i=1,2}\sum_{k} \left(G_{ki}a_k^\dagger 
	\sigma^-_i \!\!+\text{H.c.}\right)\!, \label{Hrtot2} 
\end{eqnarray}
where we assume two atoms' transition frequencies to be identical,  
and $G_{ki} = \sum_{j} g_i(x_j) G_0 e^{-i k x_{ij}}$ 
is the coupling strength
between giant atom $i$ and the waveguide. For simplicity, the
optimized coupling sequences of two atoms are assumed the same, i.e.,
$g_2(x_{i})=g_1(x_{i}+d_s)$ with $d_s$ being their separation distance.
Given that their frequencies are in the band gap [see 
Fig.~\ref{fig4m}(b)], two atoms will exchange photons without decaying. 

In principle, the exchange rate between two atoms can be tediously obtained 
by 
the
standard resolvent-operator techniques~\cite{cohen1998atom}. 
This method is valid even when 
the atom-waveguide coupling enters into the strong coupling regime. Here we 
focus on the
weak coupling regime, and the probability of photonic 
excitations in the waveguide is extremely low. In this case, 
the Rabi oscillating rate between two atoms
corresponds to their interaction strength mediated
by the waveguide’s modes, which can be simply derived
via the effective Hamiltonian method~\cite{James2007}. Only 
the modes outside the band gap interact with two atoms, and
the dipole-dipole interacting Hamiltonian mediated by one mode $k$ is derived 
as
\begin{equation}
	H_{d-d,k}=\frac{G_{k1}G^*_{k2}}{\Delta_k}(\sigma^-_1a_k^\dagger 
	\sigma^+_2a_k   
	-\sigma^+_2a_k \sigma^-_1a_k^\dagger)+\text{H.c.} \label{eff_1}
\end{equation}
The waveguide is just virtually excited and the 
photonic population is approximately zero. Therefore, by adopting the 
following
approximations
$\langle a_k^\dagger a_k \rangle\simeq0$ and $\langle a_k a_k^\dagger\rangle 
\simeq1,$
we can trace off the 
photonic freedoms in Eq.~(\ref{eff_1}), and simplify Eq.~(\ref{eff_1}) as 
\begin{equation}
	H_{d-d,k}\simeq -\frac{G_{k1}G^*_{k2}}{\Delta_k}
	\sigma^+_2 \sigma^-_1+\text{H.c.}  \label{eff_2}
\end{equation}
Note that the interaction Hamiltonian $H_{d-d,k}$ in Eq.~(\ref{eff_2}) is 
only mediated 
by one mode 
$k$. By taking all 
the modes’
contributions into account, we derive the total dipole-dipole interaction as
\begin{equation}
	H_{d-d}=J_{AB}(\sigma^+_2 \sigma^-_1+\text{H.c.}), 
\end{equation}
with 
\begin{equation}
	J_{AB}=-\sum_{k}\frac{G_{k1}G^*_{k2}}{\Delta_k}=\frac{L_w}{2\pi}\int_{-k_\text{max}}^{k_\text{max}}
	\frac{|G_{k1}|^2e^{ikd_s}}{\Delta_k}dk, \label{J12_d}
\end{equation}
where $L_w\rightarrow\infty$ is the waveguide's length adopted in the 
numerical 
simulations, and we assume that the coupling sequence of atom $b$ is 
the same with $a$, translated a distance 
$d_s$ to $a$. A long waveguide is utilized to avoid photons being reflected 
by 
the boundary. Note that we adopt the translation relation between two 
coupling sequences, i.e., 
$G_{k2}=G_{k1}e^{-ikd_s}$. From
Eq.~(\ref{J12_d}), one can find that the 
coherent exchange channel is proportional to the overlap area between two 
bound states. 
Figure~\ref{fig7m}(a) depicts $J_{AB}$ versus $d_s$
[Eq.~(\ref{J12_d})], which matches well the numerical dynamical
evolutions [obtained from the two atoms'
Rabi oscillating frequency $2J_{AB}$, see Fig.~\ref{fig7m}(b)]. 
When $\omega_q$ lies
in the band gap, due to the protection of the band 
gap 
effects, both collective and individual 
decays are zero. and the dipole-dipole exchange is free of decoherence. 

\begin{figure}[tbph]
	\centering \includegraphics[width=8.7cm]{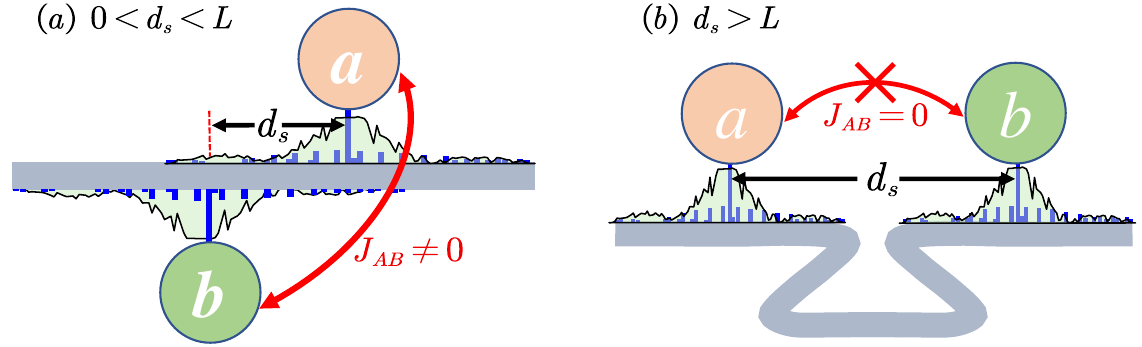}
	\caption{A QED setup where two giant atoms $a,b$ interact with 
		the same waveguide.
		(b) When two atoms' bound states overlap with each other, their
		dipole-dipole exchange rate is nonzero.
		(b) Relative to (a), two giant atoms decouple with each other when 
		their coupling 
		regimes are separated.}
	\label{fig6m}
\end{figure}

\begin{figure}[tbph]
	\centering \includegraphics[width=8.8cm]{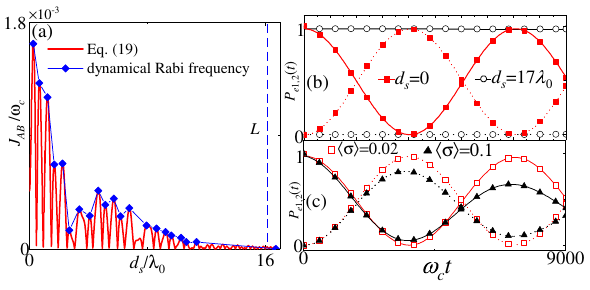}
	\caption{(a) The dipole-dipole interaction strength vs. the separation 
	distance $d_s$ 
		($\omega_q=4.4$). (b) The Rabi 
		oscillations for  $d_s=0$ and $d_s=17\lambda_0> L$, respectively. (c) 
		The dissipative 
		Rabi oscillations under different disorder 
		strengths.}
	\label{fig7m}
\end{figure} 
Since each bound state's distribution area coincide 
with the coupling regime, $J_{AB}$ is nonzero 
only when two atoms' coupling regimes overlap 
with each other. The dipole-dipole interaction vanishes
when $d_s$ is larger than the coupling distance $L$ ($L=17\lambda_0$), as depicted in
Fig.~\ref{fig6m}(b). We also consider 
two coupling sequences both experiencing independent disorders, 
the average Rabi oscillations are shown 
in Fig.~\ref{fig7m}(c). For
$\sigma_A\simeq 0.02$, the decay of the exchange process is not apparent, and the two atoms 
can 
coherently exchange excitations
with a high fidelity. 

\section{Broadband chiral emission}
In above discussions, we assume that the coupling strengths $g(x_i)$ are real 
and of the same 
sign, 
which can be realized in experiments with linear QED elements. In this case, 
the $k$-space 
interaction 
satisfies 
$G_k=G^*_{-k}$, indicating that the spontaneous emission rates into 
the right ($k>0$) and left directions 
($k<0$) are identical. Therefore, the photonic field along the waveguide has 
no 
chiral preference. Given that additional local phases are
generated via synthetic methods, i.e., 
$g(x_i)=A(x_i)e^{i\theta(x_i)}$, the relation $|G_k|=|G_{-k}|$ is not valid 
again. 
In circuit-QED this additional phase can be realized
via nonlinear Josephson 
junctions. As demonstrated in Ref.~\cite{Chen2014,Wulschner2016,Roushan2017}, 
by applying a time-oscillating flux signal with phase 
$\theta(x_i)$ through the coupling loop at point $x_i$, 
the local phase $\theta(x_i)$ were
successfully encoded into the coupling point. 
\begin{figure}[tbph]
	\centering \includegraphics[width=8.7cm]{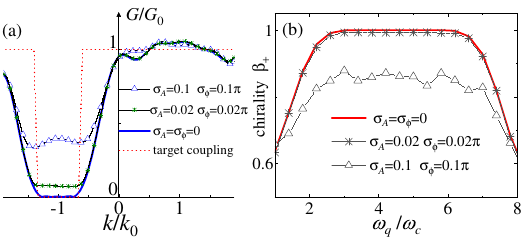}
	\caption{(a) The averaged $k$-space coupling coefficient $G_k$ for the 
		sequence in Fig.~\ref{fig1m} under different disorder strengths. (b) 
		Chiral factor 
		$\beta_+$ changes with $\omega_q$ in the presence of the disorder in 
		(a).}
	\label{fig8m}
\end{figure}

To achieve chiral emission, the left-propagating modes should be
decoupled from 
the giant atoms, a 
most important feature for $G_k$. Therefore, the 
weight function $w(k)$ inside the asymmetric gap is set to be much larger 
than outside 
the gap. Similar to realizing band-gap effects, we first 
define a 
target function:
\begin{equation}
	G^{I}_k =\begin{cases}
		0 \qquad  \left\{ -k_0-\frac{k_d}{2}<k<-k_0+\frac{k_d}{2}\right\},\\
		G_0       \qquad       \left\{-k_{\text{max}} 
		<k<-k_0-\frac{k_d}{2}\right\},  \\ 
		G_0       \qquad 
		\left\{-k_0+\frac{k_d}{2}<k<k_{\text{max}}\right\},\\
	\end{cases}
\end{equation}
where the chiral bandwidth is $k_d$.
We define a weight function during the optimizing process
\begin{equation}
	w(k)=\begin{cases}
		w_1 \qquad  \left\{ -k_0-\frac{k_d}{2}<k<-k_0+\frac{k_d}{2}\right\},\\
		w_0       \qquad       \left\{-k_{\text{max}} 
		<k<-k_0-\frac{k_d}{2}\right\},  \\
		w_0       \qquad
		\left\{-k_0+\frac{k_d}{2}<k<k_{\text{max}}\right\}.\\
	\end{cases}
\end{equation}

We optimize the target $k$-space interaction $G^I_k$ with a wide asymmetric 
band-gap (with a width $k_d/k_0=2/3$) centered at  $k_0$, as shown in 
Fig.~\ref{fig8m}(a). We 
achieve $|G_k| \neq|G_{-k}|$  
~\cite{Ramos2016,Tudela2019L,Wang2022,Chen2022}, 
indicating that chiral emission of photons can be 
observed~\cite{Caloz2018,ZhangY2021,Gheeraert2020,Kannan2022}. To 
achieve the 
optimal sets $\{x_i, A(x_i), 
\theta(x_i)\}$, we choose the constraint conditions (2-4) of 
Eq.~(\ref{constraint}). We search the optimized sequence by 
adopting
$\eta=0.125$, $N_{\text{max}}=8$, $w_1=30w_0$. and $L\simeq 2\lambda_0$.  
Similarly, searching the optimal sets $\{x_i, A(x_i), 
\theta(x_i)\}$ is now 
converted as a 
convex optimization problem by minimizing $C_m$.

Contrasting previous studies on chiral quantum
	optics targeting only on a single frequency~\cite{Mitsch2014, 
		Petersen2014}, our
	proposed system can emit photons in a broadband frequency regime, given 
	that 
	the frequency of the giant atoms is freely tuned. The amplitudes 
	$A(x_i)$ and phases $\theta(x_i)$ of the 
	coupling 
sequence are list in Table~\ref{table2} of Appendix B. In contrast to 
photonic media with 
periodic structures, the coupling 
points are now distributed with unequal spacing, and their strengths differ 
considerably. 
Moreover, 
the total
coupling number is $N=10$ and the giant atom size is $L=x_N-x_1<2\lambda_0$, 
which 
is much shorter than in a structured photonic media.
As depicted in Fig.~\ref{fig8m}(a), inside the asymmetric gap, the 
optimized $G_k$ is 
approximately zero, and matches the target coupling function. 
Compared to realizing chiral emission from a pair of 
	entangled emitters~\cite{Guimond2020}, the Lorentz reciprocity in our 
	proposed setup is
	broken. Moreover, chiral emissions in our proposal do not need any 
	preparation of 
	fragile 
	entangled states, and therefore, are more robust 
	to decoherence noise.
The chiral factor $\beta_{\pm}$ can be derived by employing the 
Weisskopf-Wigner 
theory
\begin{equation}
	\beta_{\pm}=\frac{|G_{\pm 
			k_r}|^2}{|G_{k_r}|^2+|G_{-k_r}|^2},
	\label{Grate}
\end{equation}
where $k_r=\omega_q/c$, $G_{\pm k_r}$ are the coupling 
strengths at the 
resonant 
positions, and $+(-)$ corresponds to the right (left) propagating 
mode. The asymmetric coupling with $G_{k_r}\gg 
G_{-k_r}$ indicates a right chiral emission. Moreover, the asymmetric regime 
	is very wide 
	[see Fig.~\ref{fig8m}(b)], indicating a broadband chiral emission for 
	frequency-tunable giant emitters.

When $\omega_q$ 
varies in a wide frequency regime, the chiral factor always approaches 
$\beta_+\simeq 1$.
\textit{Such broadband chiral behavior for 
	frequency-tunable giant emitters, has not yet reported in other quantum 
	setups}. For example, the chiral bandwidth of nanophotonic structures is 
	equal 
to the 
Lorentzian transmission width of the emitter, which is much narrower than 
that 
in our proposal~\cite{Coles2016}. In strongly-confined nanophotonic 
structures, the chirality is linked to spin–momentum locking, while the 
chiral emission 
in our proposal is based on interference effects. This is another 
fundamental  
advantage of our proposal.

For the experimental realization of our setup, the fabrication errors
can perturb the optimized coupling sequence.  To include this disorder 
effect, we add random 
perturbations to the coupling strength as
$g(x_i)\rightarrow [A(x_i)+\delta A(x_i)]e^{i[\theta(x_i)+\delta 
\theta(x_i)]}$ (see discussions in Appendix B).
The random offsets are sampled from Gaussian distributions with amplitude 
(phase) disorder width $\sigma_{\alpha} A(x_i)$ ($\sigma_{\phi}$).
We plot the disorder averaged $k$-space coupling for different $\{\sigma_A, 
\sigma_{\phi}\}$, as shown in 
Fig.~\ref{fig8m}(a). The asymmetric band gap is lifted  due to 
disorders. The evolution shows that the chiral factor is approximately 
one in a very wide 
frequency range for disorder strengths $\{\sigma_A=0.02, 
\sigma_{\phi}=0.02\pi\}$ [see Fig.~\ref{fig8m}(b)].
Even with stronger disorders, i.e., $\{\sigma_A=0.1, 
\sigma_{\phi}=0.1\pi\}$, the chirality remains above $0.85$, indicating that 
the 
broadband 
chiral emission realized in our proposal is robust to fabrication errors in 
the coupling 
sequence.


 \section{Conclusion}
In this work, we explore the possibilities to realize quantum optics in structured photonic 
environments with giant atoms.  We show that most phenomena can be reproduced   
by designing the couplings between giant atoms and conventional environments without any 
nanostructure. We first introduce a general  
method to find the optimized coupling sequences for arbitrary structured light-matter 
interaction. Given that a position-dependent phase is added 
to each coupling point, the giant atom can chirally emission 
photons 
in a very wide frequency regime, which has no analog in other quantum setups. We also show 
that the quantum effects in a band-gap environment (such as atomic 
fractional decay, 
static bound state and non-dissipative dipole-dipole interactions) can all be observed. 
Numerical results indicate that all the above QED phenomena can be observed even in the 
present of fabrication disorder in coupling sequences. Our proposed methods are very general 
and can also realize
other types of structured
environments, e.g., with multiple band gaps or a narrow spectrum bandwidth.  
Other quantum effects in those artificial
environments, such as non-Markovian 
dynamics or multi-photon processes, 
can also be 
revisited~\cite{Shi2016,Mahmoodian2018,Mahmoodian2020,Kusmierek22}, and new 
quantum effects might be observed.


\section{Acknowledgments} 
We thank Dr. A. F. Kockum for discussions and
useful comments. 
The quantum dynamical simulations are based on open source code 
QuTiP~\cite{Johansson12qutip,Johansson13qutip}. 
X.W.~is supported by 
the National Natural Science
Foundation of China (NSFC) (Grant No.~12174303). T.L. acknowledges the 
support from National Natural 
Science Foundation of China (Grant No.~12274142), the Startup Grant of South China University 
of Technology (Grant No.~20210012) and Introduced Innovative Team Project of Guangdong Pearl 
River Talents Program   (Grant No.~2021ZT09Z109).
F.N. is supported in part by: Nippon Telegraph and Telephone Corporation 
(NTT) Research, the Japan Science and Technology Agency (JST) [via the 
Quantum Leap Flagship Program (Q-LEAP), and the Moonshot R\&D Grant Number 
JPMJMS2061], the Asian Office of Aerospace Research and Development (AOARD) 
(via Grant No. FA2386-20-1-4069), and the Office of Naval Research (ONR) (via 
Grant No. N62909-23-1-2074).

\section*{APPENDICES}
\setcounter{equation}{0}

\setcounter{figure}{0}
\renewcommand{\theequation}{A\arabic{equation}}
\renewcommand{\thefigure}{A\arabic{figure}}

\begin{appendix}
\section{Fractional decay and bound states} 
\begin{table*}[tbp]
	\renewcommand\arraystretch{1.4}
	\begin{tabular}{>{\hfil}p{1in}<{\hfil}>{\hfil}p{0.53in}<{\hfil}>{\hfil}p{0.53in}<{\hfil}
			>{\hfil}p{0.53in}<{\hfil}>{\hfil}p{0.53in}<{\hfil}>{\hfil}p{0.53in}<{\hfil}>{\hfil}p{0.53in}<{\hfil}
			>{\hfil}p{0.53in}<{\hfil}>{\hfil}p{0.53in}<{\hfil}>{\hfil}p{0.53in}<{\hfil}>{\hfil}p{0.53in}<{\hfil}}
		\hline \hline  
		\text{position ($x_i/\lambda_0$)}& $-$8.196&	$-$7.901&	
		$-$6.992&	$-$6.682&	
		$-$4.721&	$-$4.396&	$-$3.726&	$-$3.419&$-$2.732&$-$2.441
		\\
		\hline 
		\text{coupling strength}&0.0184&	0.0291&	0.0268&	0.0146&	0.0306&	
		0.0502&	
		0.0302&	0.086&	
		0.0317 &0.1206\\
		\hline  
		\text{position ($x_i/\lambda_0$)}&	$-$1.71&	$-$1.46&	
		$-$0.507&	$-$0.006&	
		0.244&	
		0.544& 
		1.488&	2.459&3.439&	4.448
		\\
		\hline 
		\text{coupling strength}& 0.0906&	0.0748&	0.0223&	0.1413&
		0.1553&	0.9543&	0.0458&	0.1441&
		0.1305&	0.1152 \\
		\hline  
		\text{position ($x_i/\lambda_0$)}&	4.861&	5.44&	5.88&	6.383&	
		6.846&	
		7.166&	
		7.857&	8.168&&
		\\
		\hline 
		\text{coupling strength}&	0.0298&	0.0393&	0.0402&	0.0219&	0.0472&	
		0.0184&	0.0366&	0.0265 &&\\
		\hline\hline
	\end{tabular}
	\caption{The real-space coupling sequence obtained via the optimized 
		method 
		proposed in this work. The corresponding $k$-space coupling is shown 
		in 
		Fig.~\ref{fig4m}(a).} \label{table1}
\end{table*}

We first estimate the size of the decaying wavepacket for a single coupling 
point. 
According to the Weisskopf-Wigner theory, given that a small atom coupling 
at point $x_i$, the decay rate and the corresponding 
wavepacket size are respectively derived as
\begin{equation}
	\Gamma_i=\frac{2\pi|g'_{x_i}|^2}{c}, \quad 
	g'_{x_i}=g(x_i)\sqrt{\frac{L_w}{2\pi}}, \quad 
	L_i\simeq 2c\Gamma^{-1}_i.
	\label{decay_S}
\end{equation}
The coupling constant is set as $G_0=0.002\omega_c$ in our discussion. 
In numerical simulations, the mode number in the regime 
$-k_{\text{max}}<k<k_{\text{max}}$ is discretized with an interval
$\delta \! k=0.67\times10^{-3}k_0$, which is equal to 
considering a waveguide with length $L_w=1.5\times 
10^3\lambda_0$. 
Such a long waveguide guarantees the
propagating wavepacket never touches the boundary.
By employing the coupling sequence in 
Table~\ref{table1}, the maximum and average sizes of the wavepacket are 
respectively 
calculated as $\max\{L_i\}\simeq 
2\times 10^2\lambda_0$ and 
$\bar{L}_0\simeq 
8\times 10^4\lambda_0$, which are both much larger than the giant atom's size 
$L$.
Therefore, we can neglect the time retardation effects.
\begin{figure*}[tbph]
	\centering \includegraphics[width=16cm]{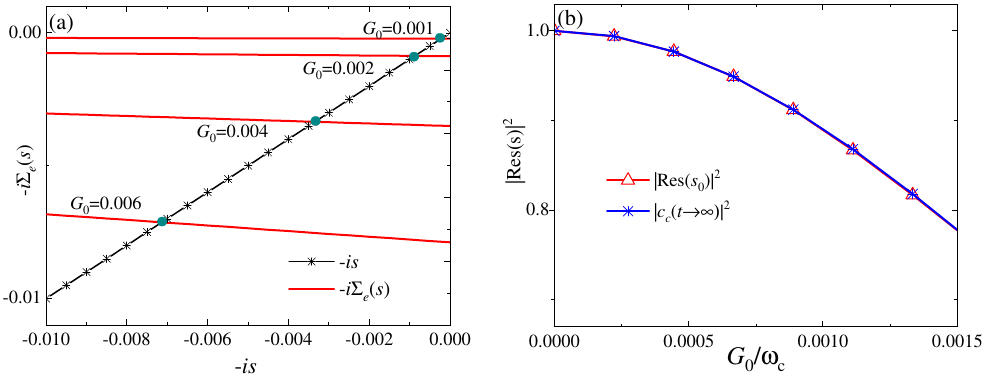}
	\caption{(a) The root of transcendental equation $s-\Sigma_e(s)=0$ can be 
		numerically solved, which 
		correspond to the intersection point (blue dots) for different values 
		of $G_0$. (b) 
		The excitation 
		population trapped inside the 
		atom $|\text{Res}(s_{0})|^2$ changes with $G_0$ [calculated via the 
		residue theorem in 
		Eq.~(\ref{RES})], and matches well the dynamical 
		evolution $|c_e(t\rightarrow\infty)|^2$. The atomic frequency is 
		fixed at 
		$\omega_q=\omega_c$.}
	\label{figA1}
\end{figure*}

Assuming a single excitation initially trapped inside the giant 
atom, the system's state at time $t$ is expanded as 
$|\psi(t)\rangle=\sum_{k} c_k(t)|g,1_k\rangle+c_{e}(t)|e,0\rangle$.
The dynamical evolution is numerically solved in this single-excitation 
subspace by discretizing the waveguide's modes in $k$ space. A similar method 
can 
be found in 
Ref.~\cite{Wang2022}. We start from the evolution governed by the interaction 
Hamiltonian 
in Eq.~(\ref{Hrtot}), which is derived as
\begin{gather}
	\dot{c}_e\left( t \right) =-i\sum_k{G_k}c_k\left( t \right) , 
	\label{Hint}	\\
	\dot{c}_k\left( t \right) =-i\Delta _kc_k\left( t \right) 
	-iG_{k}^{*}c_e\left( 
	t 
	\right).
\end{gather}
The above equations can be expressed in Laplace space as 
\begin{gather}
	s\tilde{c}_e\left( s \right) -c_e\left( t_0 \right) 
	=-i\sum_k{G_k}\tilde{c}_k\left( s \right),
	\\	
	s\tilde{c}_k\left( s \right) -c_k\left( t_0 \right) =-i\Delta 
	_k\tilde{c}_k\left( 
	s \right) -iG_{k}^{*}\tilde{c}_e\left( s \right), 
\end{gather}
and the initial conditions are 
$c_k\left(  t=0 \right) =0$ and $c_e\left( t=0 \right) =1.$
The time-dependent evolution is
derived by the inverse Laplace transformation~\cite{Ramos2016}
\begin{equation}
	c_e\left( t \right)=\frac{1}{2\pi 
		i}\lim_{E\rightarrow\infty}\int_{\epsilon-iE}^{\epsilon+iE} 
	\tilde{c}_e\left( s \right)e^{st}ds, \quad \epsilon>0.
	\label{laplace}
\end{equation}
Finally, we obtain 
\begin{equation}
	\tilde{c}_e\left( s \right)=\frac{1}{s-\Sigma_e(s)}, \quad 
	\Sigma_e(s)=\sum_k{\frac{-|G_k|^2}{s+i\Delta _k}}. \label{ces}
\end{equation}
where $\Sigma_{e}(s)$ is the self-energy of the giant atom.
Given that 
the atomic frequency is in the gap area, part of the energy will be 
trapped inside the giant atom since there is no resonant pathway to
radiate the excitation away. This point can also be verified from the roots 
of 
the transcendental 
equation $s-\Sigma_e(s)=0$, which correspond to the intersection points of
$f(s)=s$ and $f(s)=\Sigma_e(s)$ [see Fig.~\ref{figA1}(a)]. We find that 
there is only \emph{one} pure imaginary solution $s_0$ (blue 
dots), which increases with $G_0$.
Since $s_0$ is the imaginary pole for $\tilde{c}_e\left( s \right)$, 
it corresponds
to a static bound state which does not decay with 
time~\cite{cohen1998atom}. 
In this 
scenario, part of the atomic energy will be trapped without decaying, and the 
steady amplitude of $c_e(t)$ can be obtained via the residue theorem
\begin{eqnarray}
	c_e(t\rightarrow\infty)&=&\text{Res}(s_{0})=\frac{1}{1-\partial_s 
		\Sigma_{e}(s)}\Big|_{s=s_{0}} \notag \\
	&=&\frac{1}{1-\sum_k{\frac{|G_k|^2}{(s_0+i\Delta_k)^2}}}.
		 \label{RES}
\end{eqnarray}
In Fig.~\ref{figA1}(b), we plot $|c_e(t\rightarrow\infty)|^2$ versus the 
coupling strength $G_0$, which matches well with $|\text{Res}(s_{0})|^2$.
Given that the coupling strength is weak, most of 
the energy was trapped inside the atom, and the steady-state population is 
$|c_e(t\rightarrow\infty)|^2\simeq 1$. 
When increasing $G_0$, the trapped atomic excitation will decrease, and more 
energy will
distribute on the waveguide. 
\begin{figure*}[tbph]
	\centering \includegraphics[width=18.2cm]{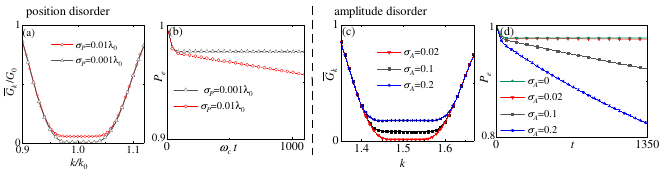}
	\caption{(a, c) The average $k$-space coupling $\bar{G}_k$ in the 
		band-gap 
		regime under different position and amplitude disorders, respectively.
		The  disorder-averaged population 
		$|\bar{p}_e(t)|$ changes with time are shown in (b, d). In each 
		realization 
		the random 
		offsets are added 
		into the optimal coupling sequence in 
		Table~\ref{table1}. Here we set $\omega_q=\omega_c$, 
		and the other parameters are the same 
		with those in Fig.~\ref{fig5m}(a).
	}
	\label{figA2}
\end{figure*}

We now show that the partial photonic 
field is trapped inside the coupling area without propagating away, which 
is akin to the bound state in QED setups with band-gap media. 
The bound state, which is the eigenstate of the system Hamiltonian, can be
obtained by solving the following
Schr{\"o}dinger
equation $H_{\text{int}}|\psi_b\rangle=E_b|\psi_b\rangle$, where $|\psi_b 
\rangle 
= \cos(\theta)|e,0\rangle + \sin \theta \sum_k c_k  \ket{g,1_k}$, with 
$\theta$ being the 
mixing angle. The solution is obtain from the following 
equations:
\begin{eqnarray}
	c_k &=& \frac{G_k}{\tan\theta(E_b - \Delta_k)}, \label{ckk} \\
	E_b &=& \sum_{k} \frac{|G_k|^2}{E_b - 
		\Delta_k}, \label{E_b} \\
	\tan \theta &=& \sum_{k} \frac{|G_k|^2}{(E_b- 
		\Delta_k)^2}.
\end{eqnarray}
Note that Eq.~(\ref{E_b}) is the same with
Eq.~(\ref{ces}) (by replacing $E_b$ with $is_0$). In our discussion, the 
interaction between the giant atom and the waveguide is weak. Therefore, the 
eigen-energy $E_b$ is around zero [i.e., $s_0\simeq 0$, see 
Fig.~\ref{figA1}(a)]. 
Under this condition, most of the energy will be trapped inside the atom, 
and the mixed angle $\theta\simeq 0$ .
Employing the approximations $\sin\theta\simeq \tan\theta$ and 
$E_b\simeq 0$, the 
photonic field is derived as 
\begin{equation}
	\psi_b(x)=\frac{\sin\theta }{\sqrt{L_w}}\sum_{k}c_ke^{kx}\simeq 
	-\frac{\sqrt{L_w}}{2\pi}\int_{-\infty}^{\infty}
	\frac{G_k}{\Delta_k}e^{kx}dk. \label{psi_bs}
\end{equation}
By substituting the real-space coupling in 
Eq.~(\ref{g_real_space}) into $\psi(x)$, we rewrite $\psi_b(x)$ as 
\begin{gather}
	\psi_b(x)=-\frac{\sqrt{L}}{2\pi}\sum_{x_i}\phi_{bi}(x), \notag \\
	\phi_{bi}(x)=\int_{-\infty}^{\infty} \frac{
		g(x_i) } {\omega_k-\omega_q}e^{k(x-x_i)}dk. 
	\label{phi_bx_i}
\end{gather}

In Eq.~(\ref{phi_bx_i}), $\phi_{bi}(x)$ is induced by a small 
atom which couples to the 1D waveguide at the single point $x_i$. 
In the weak-coupling regime, 
the small atom will exponentially decay all its energy into the waveguide 
given 
that $t\rightarrow\infty$. Therefore, there is \emph{no stable bound state} 
for a small atom, 
which can 
also be explained by the behavior 
of $\psi_{bi}(x)$ in Eq.~(\ref{phi_bx_i}), where
\begin{equation}
	\lim\limits_{\omega_k\rightarrow\omega_q}\frac{g(x_i)\, e^{-i k 
			x_i}}{\omega_k-\omega_q}=\infty \label{divergent}
\end{equation}
is divergent. That is, the expression for $\phi_{bi}(x)$ is 
non-integrable.

The counterintuitive result is that a stable bound state appears 
when all the coupling points act simultaneously. The interference between 
different points 
prevents the giant atom 
from 
decaying, and results in \emph{a static bound state even when its frequency 
lies 
	inside the 
	continuum spectrum.} 

\begin{table*}[tbp]
	\renewcommand\arraystretch{1.5}
	\begin{tabular}{>{\hfil}p{1.2in}<{\hfil}>{\hfil}p{0.5in}<{\hfil}>{\hfil}p{0.5in}<{\hfil}
			>{\hfil}p{0.5in}<{\hfil}>{\hfil}p{0.5in}<{\hfil}>{\hfil}p{0.5in}<{\hfil}>{\hfil}p{0.5in}<{\hfil}
			>{\hfil}p{0.5in}<{\hfil}>{\hfil}p{0.5in}<{\hfil}>{\hfil}p{0.5in}<{\hfil}>{\hfil}p{0.5in}<{\hfil}}
		\hline \hline  
		\text{position} ($x_i/\lambda_0$) 
		&$-$0.909&-0.757&$-$0.383&$-$0.508&$-$0.0975&$-$0.222&0.393
		&0.120&0.641&0.909
		\\
		\hline 
		\text{amplitude} $A(x_i)$ & 0.088&0.130&0.628&0.429&0.392&0.591
		&0.365&0.198&0.615&0.243\\
		\hline  
		\text{phase} $\theta(x_i)$ 
		&0.388$\pi$&$-$0.500$\pi$&$-$0.446$\pi$&0.500$\pi$&$-$0.500$\pi$
		&0.500$\pi$&$-$0.500$\pi$&0.179$\pi$&0.0048$\pi$&0.460$\pi$\\
		\hline\hline
	\end{tabular}
	\caption{The amplitudes and phases of the coupling sequence in 
	Fig.~\ref{fig7m}, which is obtained via the proposed optimization 
		method.} \label{table2}
\end{table*}
\setcounter{equation}{0}

\setcounter{figure}{0}
\renewcommand{\theequation}{B\arabic{equation}}
\renewcommand{\thesubsection}{B\arabic{subsection}}
\renewcommand{\thefigure}{B\arabic{figure}} 
\section{Simulating disorder effects }
\subsection{Disorder effects in band-gap environments}
When implementing the optimal coupling sequences in experiments, there 
will be 
fabrication errors in both coupling positions and amplitudes. Next we 
evaluate 
their effects by considering circuit-QED with a transmon qubit 
(with frequency $\omega_q/(2\pi)=3~\text{GHz}$, for example). The phase 
velocity 
along the transmission-line waveguide is set as $c=2\times 
10^8\text{m}/\text{s}$. 
Therefore, the wavelength is $\lambda_0\simeq 7\times 10^{-2}\text{m}$. We 
first 
investigate the disorder in positions by adding random offsets to the 
coupling 
sequence, i.e., 
$g(x_i)\rightarrow g(x_i+\delta x_i)$. Here $\delta x_i$ is sampled from 
a 
Gaussian distribution centered around zero and with a width $\sigma_{P}$.

\begin{figure*}[tbph]
	\centering \includegraphics[width=18cm]{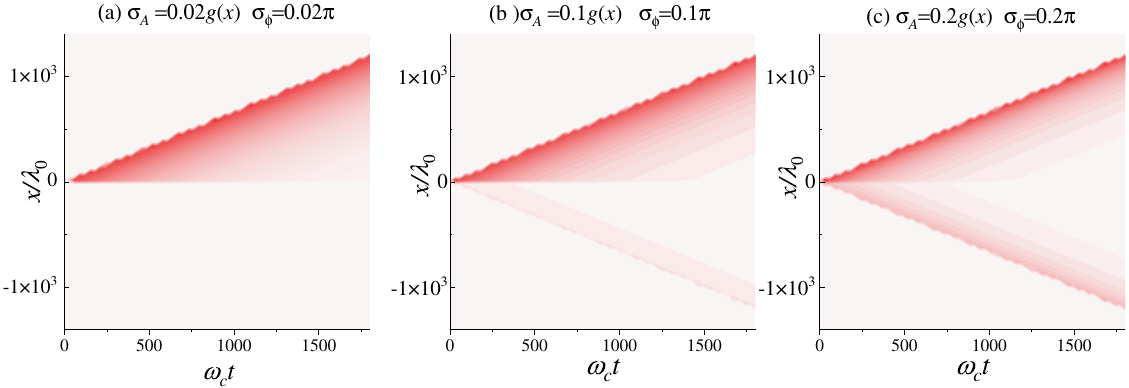}
	\caption{(a-c) Time evolution 
		of the chiral 
		field distributions for various disorder strengths. The atomic 
		frequency is fixed at 
		$\omega_q=\omega_c$.
	}
	\label{figA3}
\end{figure*}

Consequently, the average $k$-space 
coupling function is defined as
\begin{equation}
	\bar{G}_k=\frac{1}{N_{\text{dis}}}\sum_{n=1}^{N_{\text{dis}}} 
	\sum_{x_i}{g(x_i+\delta x_i)
		e^{-ik(x_i+\delta x_i)}},
\end{equation}
where $N_{\text{dis}}$ is the number of disorder realizations in the 
numerical 
simulations. 
In our discussion, we set $N_{\text{dis}}=200$, which is large enough for the 
errors 
considered in this work. To investigate the disorder effects on the quantum 
dynamics, 
we numerically simulate the evolution by taking the average of all the 
realizations. We define the disorder-averaged
population as
\begin{equation}
	\bar{p}_e(t)=\frac{1}{N_{\text{dis}}}\sum_{n=1}^{N_{\text{dis}}}\left|c_e(t)\right|^2.
\end{equation}

In Fig.~\ref{figA2}(a), we plot $\bar{G}_k$ 
for different disorder strengths, and find that the band gap
is lifted higher than zero. Given that $\sigma_{P}=0.001\lambda_0\simeq 
0.07~\text{mm}$, the coupling strength in the band gap is  
around zero, and the giant atom is still protected from decaying [see 
Fig.~\ref{figA2}(b)]. Only when the 
location error is $\sigma_{P}>0.01\lambda_0\simeq 0.7~\text{mm}$ (which 
should 
already be visible to the naked eye), the band gap will be lifted around 
$\bar{G}_k\simeq 0.05$ and the transmon will gradually decay due to
the position
disorders. Current and future experiments on circuit-QED can locate 
coupling elements with accuracy higher than $0.01\lambda_0$. 
Therefore, we can neglect the position errors in our discussion.  

Next we consider random offsets to the amplitude coupling sequence, i.e., 
$g(x_i)\rightarrow g(x_i)+\delta g(x_i)$. Here $\delta g(x_i)$ is sampled 
from 
a 
Gaussian distribution centered around zero and with a width $\sigma_A\,
g(x_i)$. In Fig.~\ref{figA2}(c), we plot $\bar{G}_k$ 
for different disorder strengths, and find that the band gap
is lifted higher than zero. 
The trapped excitation inside the atom becomes unstable and 
will slowly leak into the waveguide [see Fig.~\ref{figA2}(d)]. The 
decoherence rate led by disorder increases with disorder strengths.
It can be inferred that when $\sigma_A>0.2$, the protection effects of the 
band 
gap
will be swamped by the disorder noise. 

\subsection{Disorder effects in broad-band chiral emission}
The obtained optimal
set $\{x_i, A(x_i), 
\theta(x_i)\}$ for broad-band chiral emission are listed in 
Table~\ref{table2}, which is plotted in 
Fig.~\ref{fig1m}. We now consider that the coupling strength at each point 
experiences 
disorders in both 
its amplitude and phase, i.e., 
$g(x_i)\rightarrow [A(x_i)+\delta A(x_i)]\exp[{i\theta(x_i)+i\delta 
\theta(x_i)}].$
Both the amplitude and phase disorders are assumed to satisfy a Gaussian 
distribution centered 
around 
zero. The amplitude disorder widths $\sigma_A$ are proportional to the local 
strength $A(x_i)$, 
while the phase disorder widths $\sigma_{\phi}$ are assumed identical for all 
the coupling 
points. We plot the disorder averaged $k$-space coupling function 
in 
Fig.~\ref{fig8m}(a). We find that the disorder does not 
affect the coupling strength too much for the modes outside the chiral 
regime. Inside the asymmetric band gap, the zero coupling will be 
lifted higher than zero with stronger disorder strengths. 

To show how disorder disturbs the chiral emission, we numerically simulate 
disorder-averaged
evolutions by defining the photonic field as
\begin{eqnarray}
	\bar{\Psi}_{\gamma}(x,t)&=&\frac{1}{N_{\text{dis}}}\sum_{n=1}^{N_{\text{dis.}}}
	|\psi_{\gamma}(x,t)|^2, \\
	\psi_{\gamma}(x,t)&\propto& \int_{-\infty}^{\infty} dk c_k(t)e^{-ikx}.
	\label{field_D}
\end{eqnarray}

In Fig.~\ref{figA3}(a-c), we plot how the disorder-averaged field 
distribution 
$\psi_{\gamma}(x,t)$ changes with time in the present of $\{\sigma_A, 
\sigma_{\phi}\}$. When the coupling disorders are as strong as 
$\{\sigma_A=0.1g(x_i), 
\sigma_{\phi}=0.1\pi\}$, most of the 
photonic field still decays to the right of the waveguide. 
To evaluate the chiral behavior of 
our proposal, we define the chiral factor as
\begin{gather}
	\beta_\pm=\frac{\Phi_{R(L)}}{\Phi_{R}+\Phi_{L}}, \\
	\Phi_{R/L} = \frac{1}{N_{\text{dis}}}\sum_{\text{dis.}}\lim_{t\rightarrow 
	\infty}\left| 
	\int_{0}^{\pm \infty} |\psi_{\gamma}(x',t)|^2 
	dx' \right|, 
	\label{chiral_fac}
\end{gather}
Employing the above methods and definitions,
we plot Fig.~\ref{fig8m}, which shows that our 
proposal can chirally route photons in a broadband range even in the present 
of 
strong disorder.

There are many types of layouts which can encode the required phases via 
nonlinear couplings. For 
example, by applying a time-dependent flux through a coupler loop at position 
$x_i$, 
the coupling strength can be written as 
\begin{equation}
	g(x_i,t)=g_i\frac{\Phi_{\text{ext}}}{\Phi_{0}}\cos(\Omega_d 
	t+\phi_i),
\end{equation}
where $g_i$ is the coupling constant depending on the circuit 
parameters [such as the Josephson inductance and loop inductance],
$\Phi_{\text{ext}}$ 
($\Omega_d$) is the time-dependent driving amplitude (frequency), and 
$\phi_{i}$ is 
the phase to be encoded at $x_i$. 
Therefore, 
we do not require the circuit parameters to be fabricated to a certain value. 
If 
$g_i$ is smaller (larger) than the required value, the external 
driving amplitude $\Phi_{\text{ext}}$ can be increased 
(decreased) accordingly when calibrating the whole 
setup. Therefore, the disorder is affected by the drive, which 
is usually the
output from devices such as arbitrary function generators [see e.g., 
\href{https://www.tek.com/en/products/signal-generators}{https://www.tek.com/en/products/signal-generators}].
In many labs, both 
the amplitude and phase of the drive can be controlled with high 
accuracy, 
indicating that the disorder in the nonlinear coupling layout can be 
suppressed to 
low values in experiments. 
	
\end{appendix}	
	
%


\end{document}